\documentclass[conference]{IEEEtran}
\IEEEoverridecommandlockouts
\usepackage{cite}
\usepackage{amsmath,amssymb,amsfonts,amsthm}

\usepackage{algorithmic}
\usepackage{graphicx}
\usepackage{textcomp}
\usepackage{xcolor}
\usepackage{booktabs}
\usepackage{xspace}
\usepackage{hyperref}
\def\BibTeX{{\rm B\kern-.05em{\sc i\kern-.025em b}\kern-.08em
    T\kern-.1667em\lower.7ex\hbox{E}\kern-.125emX}}

\newcommand{\transp}{\ensuremath{^\mathsf{T}}\xspace}
\newcommand{\figref}[1]{Fig.~\ref{#1}\xspace}

\begin{document}

\title{Robust Zonotopic Control}

\author{\IEEEauthorblockN{Fouzi Tabouri}
\IEEEauthorblockA{%
\textit{Aalborg University}\\
Aalborg, Denmark \\
fota@cs.aau.dk}
\and
\IEEEauthorblockN{Kim Guldstrand Larsen}
\IEEEauthorblockA{%
\textit{Aalborg University}\\
Aalborg, Denmark \\
kgl@cs.aau.dk}
\and
\IEEEauthorblockN{Christian Schilling}
\IEEEauthorblockA{%
\textit{Aalborg University}\\
Aalborg, Denmark \\
christianms@cs.aau.dk}}

\maketitle

\begin{abstract}
We propose a zonotopic framework for synthesizing a single robust state feedback controller that is certified to stabilize every plant inside a matrix zonotope, describing linearly varying parameters or parametric uncertainty. Common robust design strategies rely on checking many vertex models or on complex gain-scheduling, leading to high offline computation and implementation complexity. Our approach finds a single gain that is provably valid across the entire parameter domain, which is simpler to implement and can reduce conservatism by exploiting the structure of the zonotope. We formulate the robust synthesis as a single convex program tailored to the zonotope representation and incorporate practical performance requirements (actuator constraints, decay rate, disturbance attenuation) into the same synthesis stage. In numerical experiments on a representative 4-state example, our controller provides larger stability coverage across the parameter domain, attains comparable transient performance and control effort to more complex designs, and significantly reduces the number and scale of offline synthesis problems required by other robust approaches, compared to common-vertex gain, \(H_{\infty}\), and \(\mu\)-synthesis baselines.
\end{abstract}


\section{Introduction}

Modern safety-critical control tasks face varying conditions and sizable uncertainties that can compromise stability and performance. Linear Parameter Varying (LPV) representations provide structured, tractable, parameter-dependent models, including bounds and parameter-variation rates when relevant~\cite{zhou_lpv,toth2010modeling,wu1996induced}. A central challenge is to represent the operating/uncertainty set so it captures coupled/correlated variations and allows for efficient certification and controller design. Common practical choices include polytopic, ellipsoidal, or norm-bounded sets~\cite{mohammadpour2012control,toth2010modeling}.
In this paper, we propose to use zonotopes, a compact, generator-based set representation, to express correlated parameter variations.

\subsection{Related Work}

Linear matrix inequalities (LMIs)
are the basis of modern \(H_{\infty}\) and robust control synthesis~\cite{boyd1994linear,APKARIAN19951251,chilali2002h}. The S-procedure~\cite{yakubovich1971sprocedure} is the standard tool to convert quadratic constraints into tractable LMI certificates, able to translate many `for all' requirements into finite semi-definite conditions. Robust control for LPV systems has been extensively studied for their ability to model parametric uncertainties and varying operational conditions~\cite{zhou_lpv,toth2010modeling,wu1996induced}. In polytopic LPV representations, the plant is formulated as a convex combination of vertex models (often representing worst cases). The LMIs are expressed at those vertices as the classical approach for gain-scheduling and LPV control design and synthesis. The method gives exact finite tests (vertex checks) at the cost of an exponential vertex count in the number of varying/uncertain parameters~\cite{sename2025linear,mohammadpour2012control}.

Zonotopes~\cite{le2013zonotopes} are common in reachability, tube-MPC, and set-membership estimation~\cite{girard2005reachability,althoff2010reachability,althoff2011analyzing,le2013zonotopic,luo2023reachability,alanwar2021data}, and have been applied to LPV state estimation and fault detection~\cite{Set_based_FDI_LPV,Zono_Set_Switched_LPV}, and robust MPC to ensure constraint satisfaction under uncertainty and improve state estimation~\cite{le2011robust,le2013zonotopic}, or enhance the optimization process computational cost~\cite{andrade2024tube}. The zonotopic representation has been extended to matrices (matrix zonotopes) to represent sets of system matrices and compute reachable sets in the case of uncertain time-varying systems~\cite{AlthoffGK11,althoff2011analyzing,luo2023reachability,alanwar2021data}. Few recent works explicitly leverage zonotopic uncertainty and operating-domain variations in controller design for uncertain switched systems~\cite{zhang2024linear} and motion planning~\cite{carrizosa2023zonotopic}.

To summarize, LMI methods and polytopic LPV synthesis provide certifiable ways to design controllers for varying operational conditions/parametric uncertainties, but they suffer from exponential vertex growth if the number of varying parameters increases. Zonotopes offer a convenient description of operating sets and have been used successfully for control-related goals. However, zonotopic representations have not yet been employed to directly design a single robust control via a single convex program; this is the goal of this paper.

\subsection{Motivation and Contributions}

Our work is motivated by the powerful but expensive vertex method for controller synthesis. Combining LMI methods, S-procedure, and the compactness of zonotopic representations, we propose a synthesis method that directly constructs a single robust state feedback controller that is certified for all plants inside a zonotope describing the parameter/operating domain. The zonotopic constraints are cast into one block LMI with no need for vertex enumeration, is convex in the decision variables, and is adaptable to performance specifications like decay rate and \(H_{\infty}\) performance weighting.
Thus, robust zonotopic control is a compact, stable approach for many LPV problems. Our key contributions are:
\begin{itemize}
    \item We design a single tractable block LMI that certifies closed-loop stability for all plants inside the zonotope.
    \item Our framework allows for efficient synthesis, reducing the complexity over the vertex-based baseline from exponential to polynomial, without additional assumptions.
    \item Our approach demonstrates state-of-the-art performance against vertex-based, \(H_{\infty}\), and \(\mu\)-synthesis controllers across rigorous benchmarks.
    \item We discuss conservatism, sufficiency and implementation.
\end{itemize}

\section{Foundational Concepts}
\label{sec:lpv_sys}
In this section, we introduce LPV systems, zonotopes, and the zonotopic representation of varying parameters.

\subsection{LPV Systems}
The state-space equations of continuous LPV systems are
\begin{align}
    \dot{x}(t) &= A(\rho(t))x(t)+B(\rho(t))u(t), \label{eq:lti_orig} \\
    y(t) &= C(\rho(t))x(t)+D(\rho(t))u(t), \nonumber
\end{align}
where \(\rho(t) = (\rho_1(t), \rho_2(t), \dots,\rho_N(t))\) is the vector of varying parameters that evolves inside a known compact set. One may consider \(\rho(t)\) as either time-varying parametric uncertainty or the combination of measurable signals indicating the operating point at time~\(t\).
We uniformly refer to the operating domain and the uncertainty set as the varying parameter set~$\Omega$.
Using the pre-filtering trick on the control input \(u\)~\cite{APKARIAN19951251}, we may drop the parameter dependence of \(B\) in Eq.~\eqref{eq:lti_orig}.
We also assume that \(A\) has affine dependence on \(\rho(t)\).
Thus, Eq.~\eqref{eq:lti_orig} becomes
\begin{align}
    &\dot{x}(t) = A(\rho(t))x(t)+Bu(t) \label{eq:lti} \\
    \text{where } \quad &A(\rho(t)) =A_0+\sum_{i=1}^N\rho_iA_i.
    \label{eq:affine_rho}
\end{align}

Moreover, we assume that the following common conditions on the varying parameters \(\rho(t)\) are always satisfied~\cite{SenameBook}:
\begin{itemize}
    \item The parameter domain \(\Omega \ni \rho_i\) is compact and convex.
    \item The time-varying matrices $A$ are continuous in \(\Omega\).
    \item The parameters \(\rho_i\) are exogenous (not state dependent).
    \item For parameter-varying (in contrast to uncertain) models, the parameters \(\rho_i\) are measurable at every instant~\(t\).
\end{itemize}

\subsection{Zonotopes}
A zonotope \(Z \subseteq \mathbb{R}^N\) is an affine map of the unit hypercube,
\begin{equation*}
    Z = \left\{ c+\sum_{i=1}^mg_id_i \mid d_i\in[-1,1], i=1,\dots,m \right\},
\end{equation*}
with center \(c \in \mathbb{R}^N\) and $m$ generators \(g_i \in \mathbb{R}^N\).
Zonotopes are centrally symmetric polytopes.

\subsection{Zonotopic Representation of the Varying Parameter Set}

In robust control for LPV systems, the varying parameters \(\rho_i\) are often independent intervals. For $N$ varying parameters, this yields a hyperrectangular set~$\Omega$ of \(2^N\) vertices.
Thus, the parameter-dependent system matrix~$A$ at any time is a convex combination of the matrices at these vertices:
\begin{equation*}
    A(\rho(t))=\sum_{i=1}^N\alpha_i(\rho(t))A_i
\end{equation*}
where \(\alpha_i(\rho(t))\geq0\), \(\sum_{i=1}^N\alpha_i(\rho(t))=1\), and \(A_i\) is a constant matrix obtained by expressing \(A(\rho(t))\) at the vertex \(i\).

In this work, we represent~$\Omega$ as a zonotope instead of the traditional vertex-based polytope,
as follows:
\begin{equation*}
    \Omega = \left\{ \rho_c+\sum_{i=1}^mg_id_i \mid d_i\in[-1,1],\quad \forall i \in\{1,\dots,m\} \right\}
\end{equation*}

Any point \(\rho=(\rho_1, \rho_2, \dots, \rho_N)\transp\) in this set is expressible as a combination of
the generators \(g_i=[g_{i,1}, g_{i,2}, \dots,g_{i,m}]\transp\).
Thus,
the component-wise expression of the parameter \(\rho_i\) is
\begin{equation}\label{eq:rho_di}
    \rho_i=\rho_{c,i}+\sum_{j=1}^mg_{j,i}d_j.
\end{equation}

This alternative representation allows us to reason about the zonotopic coefficients \(d\) instead of the varying parameters \(\rho\).
Note that for hyperrectangular~$\Omega$, $m = N$.

\section{Control Problem and Solution}
\label{sec:control_pb}
In this section, we formulate the control problem, present our robust zonotopic control framework, and discuss extensions to performance specifications.

\subsection{Zonotopic Formulation of the System}
\label{sec:zono_form}
We can rewrite the matrix~\(A(\rho(t))\) using the affine dependence on \(\rho(t)\) (Eq.~\eqref{eq:affine_rho}) together with Eq.~\eqref{eq:rho_di}:
\begin{equation*}
\begin{aligned}
A(\rho(t)) &= A_0 + \sum_{i=1}^N \rho_i A_i
= A_0 + \sum_{i=1}^N (\rho_{c,i} + \sum_{j=1}^m g_{j,i} d_j) A_i \\
&= A_0 + \sum_{i=1}^N \rho_{c,i} A_i + \sum_{i=1}^N \sum_{j=1}^m g_{j,i} d_j A_i \\
&= \underbrace{A_0 + \sum_{i=1}^N \rho_{c,i} A_i}_{=:\,\widetilde{A_0}} + \sum_{j=1}^m d_j \underbrace{\Big(\sum_{i=1}^N g_{j,i} A_i\Big)}_{=:\,\widetilde{A_j}} \\
\end{aligned}
\end{equation*}

Therefore, system~\eqref{eq:lti} can be written as
\begin{equation*}
    \dot{x}(t) = A(d)x(t)+Bu(t),\quad d_i\in[-1,1],
\end{equation*}
where \(A(d)=\widetilde{A_0} + \sum_{j=1}^m d_j \widetilde{A_j}\) parameterizes a matrix zonotope~\cite{AlthoffGK11}. We explicitly treat \(A(d)\) as a matrix-valued function for a fixed \(d \in [-1,1]^m\), distinct from the set-valued view. 

\subsection{Zonotopic Robust Control Framework}
\label{sec:zono_frame}
We aim to design a state feedback controller \(u=Kx\) that robustly stabilizes the closed-loop system for all admissible values of \(d\) across the given zonotope. Having expressed the open-loop system in zonotopic form as well, we get
\begin{equation}\label{eq:CLS_eq}
    \dot{x}(t)
    =A(d)x(t)+Bu(t)
    =(A(d)+BK)x(t).
\end{equation}

For a stabilizing controller~\(K\),
the closed-loop system must admit a quadratic Lyapunov function \(V(x)=x\transp Px,\)
with \(P\) symmetric positive definite (i.e., \(P=P\transp >0\)) such that
\begin{equation}\label{eq:vdot_ineq}
    \dot{V}(x)=\dot{x}\transp Px+x\transp P\dot{x}<0, \quad \forall x\neq0.
\end{equation}

Using Eq.~\eqref{eq:CLS_eq} to rewrite the condition in Eq.~\eqref{eq:vdot_ineq}, and letting
\(\Phi(d):=(\widetilde{A_0}+BK)\transp P+P(\widetilde{A_0}+BK)+\sum_id_i(\widetilde{A_i}\transp P+P\widetilde{A_i})\),
we obtain (for all $x\neq 0$ and for $d_i\in[-1,1]$ for all $i$):
\begin{align*}
    \dot{V}(x)= x\transp \Phi(d) x<0 \iff \Phi(d)<0
\end{align*}

Let \(Q:=P^{-1}\) and \(Y:=KQ\). Since \(P=P\transp >0\), we have \(Q=Q\transp >0\); thus, \(\Phi(d)<0 \iff Q\transp \Phi(d)Q<0 \iff Q\Phi(d)Q<0\),
and we can write~\eqref{eq:vdot_ineq} as
\begin{equation*}
\begin{aligned}
    & \dot{V}(x)<0\iff Q\Phi(d)Q<0 \\
    \!\iff & \underbrace{(\widetilde{A_0}Q \!+\! BY) \!+\! (\widetilde{A_0}Q \!+\! BY)\transp}_{=: M_0} \!+\! \sum_i \! d_i \underbrace{(Q\widetilde{A_i}\transp \!\!+\!\widetilde{A_i}Q)}_{=: M_i} \!<\! 0.
\end{aligned}
\end{equation*}
We can write the constraints on \(d_i\) in quadratic and LMI form:
\begin{equation}\label{eq:quad_mat_di}
    \|d_i\|\leq1
    \iff (1-d_i^2)\geq0
    \overset{\forall x\neq0}{\iff} x\transp (1-d_i^2)I_nx\geq0,
\end{equation}
with \(I_n\) the identity matrix (of appropriate order~$n$).

We want that \eqref{eq:quad_mat_di}\(\implies\)(\(M_0+\sum_id_iM_i<0 \quad \forall d_i\in[-1,1]\)). By applying the S-procedure, a sufficient condition is
\begin{align*}
    \exists\mu_i\geq0 & \text{ s.t. } -[M_0+\sum_id_iM_i]>\sum_i\mu_i(1-d_i^2)I_n \\
    \iff& x\transp [M_0+\sum_i\mu_iI_n+\sum_id_iM_i-\sum_i\mu_id_i^2I_n]x<0,
\end{align*}
which can be cast into a single full block LMI as follows:
\begin{equation*}
\begin{aligned}
& x\transp \Big[M_0+\sum_i \mu_i I_n+\sum_i d_i M_i-\sum_i \mu_i d_i^2 I_n\Big]x \\
={} & x\transp (M_0+\sum_i \mu_i I_n)x + \sum_i (d_i x)\transp \Big(\tfrac{1}{2} M_i\transp \Big) x
+ {} \\
& \sum_i x\transp \Big(\tfrac{1}{2} M_i\Big)(d_i x) - \sum_i (d_i x)\transp  \mu_i I_n (d_i x) < 0
\end{aligned}
\end{equation*}

Defining $\mathcal{L}(Q,Y,\mu)$ as
\begin{equation*}
    \begin{bmatrix}
        M_0+\sum\limits_i\mu_iI_n & \frac{1}{2}M_1 & \frac{1}{2}M_2 & \cdots & \frac{1}{2}M_m\\
        \frac{1}{2}M_1\transp  & -\mu_1I_n & 0_{n\text{x}n} & \cdots & 0_{n\text{x}n} \\
        \frac{1}{2}M_2\transp  & 0_{n\text{x}n} & -\mu_2I_n & \cdots & 0_{n\text{x}n} \\
        \vdots & \vdots & \vdots & \ddots & \vdots \\
        \frac{1}{2}M_m\transp  & 0_{n\text{x}n} & 0_{n\text{x}n} & \cdots & -\mu_m I_n
    \end{bmatrix}
\end{equation*}
and \(z\transp  \!:=\! \begin{bmatrix}x\transp  & (d_1x)\transp  &  (d_2x)\transp  & \cdots & (d_mx)\transp \end{bmatrix}\), we get:
\begin{equation*}
    z\transp     \mathcal{L}(Q,Y,\mu)
    z
    < 0
     ,\quad \forall z\neq0
\end{equation*}
This corresponds to the LMI $\mathcal{L}(Q,Y,\mu)<0$. Thus, if such \(Q=Q\transp >0\), \(Y\), and scalars \(\mu_i\geq0\) exist, the state feedback gain \(K=YQ^{-1}\) renders \(A(d)+BK\) exponentially stable for any admissible \(d\) with \(\|d_i\|\leq1\), and hence for any \(\rho(t) \in \Omega\). Note that the S-procedure test is sufficient but not necessary: stabilizing gains $K$ may exist for which the scalar multipliers \(\mu_i\) do not certify. To increase precision, one can replace the multipliers \(\mu_iI_n\) with positive definite Hermitian matrices (\(\mathrm{T}_i\)), as employing a single positive scalar multiplier \(\mu_i\) multiplied by the identity matrix imposes more restrictive constraints, and allowing matrix-valued multipliers relaxes them:
\begin{align*}
    ||d_i||\leq1
    &\iff d_i^2\leq 1
    \iff d_i^2\mathrm{T}_i\leq\mathrm{T}_i\\
    &\iff z^\text{T}(d_i^2\mathrm{T}_i)z\leq z^\text{T}\mathrm{T}_iz \\
    &\iff z^\text{T}\mathrm{T}_iz-(d_iz)^\text{T}\mathrm{T}_i(d_iz)\geq0, \quad \forall z\neq 0
\end{align*}

The corresponding LMI $\mathcal{L}(Q,Y,\mathrm{T}_i)$ thus is
\begin{equation*}
    \begin{bmatrix}
        M_0+\sum_i\mathrm{T}_i & \frac{1}{2}M_1 & \frac{1}{2}M_2 & \cdots & \frac{1}{2}M_m\\
        \frac{1}{2}M_1\transp  & -\mathrm{T}_1 & 0_{n\text{x}n} & \cdots & 0_{n\text{x}n} \\
        \frac{1}{2}M_2\transp  & 0_{n\text{x}n} & -\mathrm{T}_2 & \cdots & 0_{n\text{x}n} \\
        \vdots & \vdots & \vdots & \ddots & \vdots \\
        \frac{1}{2}M_m\transp  & 0_{n\text{x}n} & 0_{n\text{x}n} & \cdots & -\mathrm{T}_m
    \end{bmatrix}<0
\end{equation*}

\subsection{Performance Specifications}
\label{sec:perf_spec}
So far, the zonotopic LMI formulation only proves stability.
Now we add performance specifications for addressing input constraints related to actuator saturation and decay rate.

\subsubsection{Input constraints: sector-bounded nonlinearities}
\label{sec:sectorbound}
Nonlinearities that affect actuator saturation can be represented as sector-bounded functions.
This allows to include their effect in the zonotopic LMI formulation
(Section~\ref{sec:zono_frame}) using the S-procedure, guaranteeing stability across the zonotopic set.

Here we study saturation nonlinearity with a linear region
\begin{equation*}
    u = v - \phi(v),
\end{equation*}
where \(v = Kx\) is the linear component (controller output) and \(\phi(v)\) is the nonlinear component (saturation effect), satisfying the sector condition \( \phi(v) \in \text{sector}[0,1]\), which we write as
\begin{equation*}
    \phi(v)\transp (\phi(v)-v)\le0,
\end{equation*}
(see~\cite{khalil2002nonlinear}).
The closed-loop dynamics, within the context of robust zonotopic state feedback control~\eqref{eq:CLS_eq}, is now written as
\begin{align*}
    \dot{x}
    = A(d)x+Bu
    = (A(d)+BK)x-Bw,
\end{align*}
with \(w=\phi(Kx)\). Rewriting~\eqref{eq:vdot_ineq} including the sector condition:
\begin{align*}
    \dot{V}(x,w)
    ={}& x\transp (A\transp (d)+K\transp B\transp )Px+x\transp P(A(d)+BK)x \\
    & -w\transp B\transp Px-x\transp PBw <0
\intertext{Let \(Q:=P^{-1}\), \(z:=Px\), and \(Y:=KQ\), and recall that $P = P\transp $ and $Q = Q\transp $.}
    ={} & z\transp (QA_0\transp +A_0Q+Y\transp B\transp +BY +{} \\
    & \sum_id_i(QA_i\transp +A_iQ))z -w\transp B\transp z-z\transp Bw<0
\end{align*}
We can reformulate the sector condition as:
\begin{align*}
    w\transp (w-Kx)\le0& \!\iff\! w\transp (\frac{1}{2}Y)z+z\transp (\frac{1}{2}Y\transp )w-w\transp w\geq0
\end{align*}

\begin{figure*}[t]
  \begin{minipage}{\textwidth}
  \begin{equation}
    \label{eq:dTMd_performance}
    \resizebox{.958\columnwidth}{!}{$%
    \hspace*{-2.5mm}
    \begin{bmatrix}
        z\\d_1z\\d_2z\\ \vdots \\d_mz\\w
    \end{bmatrix}\transp 
    \left[
    \begin{array}{ccccc|c}
        QA_0\transp +A_0Q+Y\transp B\transp +BY+\sum_i\mu_iI_n & \frac{1}{2}(QA_1\transp +A_1Q) & \frac{1}{2}(QA_2\transp +A_2Q) & \cdots & \frac{1}{2}(QA_m\transp +A_mQ) & -B+\lambda\frac{1}{2}Y\transp \\
        \frac{1}{2}(QA_1\transp +A_1Q)\transp  & -\mu_1I_n & 0_{n\text{x}n} & \cdots & 0_{n\text{x}n} & 0_{n\text{x}1}\\
        \frac{1}{2}(QA_2\transp +A_2Q)\transp  & 0_{n\text{x}n} & -\mu_2I_n & \cdots & 0_{n\text{x}n} & 0_{n\text{x}1}\\
        \vdots & \vdots & \vdots & \ddots & \vdots & \vdots \\
        \frac{1}{2}(QA_m\transp +A_mQ)\transp  & 0_{n\text{x}n} & 0_{n\text{x}n} & \cdots & -\mu_m I_n & 0_{n\text{x}1}\\
        \hline
        -B\transp +\lambda\frac{1}{2}Y & 0_{1\text{x}n} & 0_{1\text{x}n} & \cdots & 0_{1\text{x}n} & -\lambda
    \end{array}
    \right]
    \begin{bmatrix}
        z\\d_1z\\d_2z\\ \vdots \\d_mz\\w
    \end{bmatrix}<0
    $}
  \end{equation}
  \hrule
  \end{minipage}
\end{figure*}

We now write the constraints on \(d_i\):
\begin{align*}
    d_i^2\leq1
    \iff d_i^2z\transp z\leq z\transp z
    \iff z\transp z-(d_iz)\transp (d_iz)\geq0
\end{align*}

Using the S-procedure, a sufficient condition to guarantee stability over the zonotopic set with the sector-bounded nonlinearity is to find positive multipliers \(\mu_i\) and \(\lambda\) such that
\begin{align*}
    \dot{V} + {} & \sum_i\mu_i(z\transp z-(d_iz)\transp (d_iz)) + {} \\
    & \lambda(w\transp (\frac{1}{2}Y)z+z\transp (\frac{1}{2}Y\transp )w-w\transp w)<0,
\end{align*}
which can be cast into a single full block LMI as follows:
\begin{equation*}
\begin{aligned}
    z\transp  & (QA_0\transp  \!+\! A_0Q \!+\! Y\transp B\transp  \!+\! BY)z \!+\! \sum_i(d_iz\transp (QA_i\transp  \!+\! A_iQ)z) \\
    & -w\transp B\transp z-z\transp Bw
    +\sum_i\mu_i(z\transp z-(d_iz)\transp (d_iz)) \\
    & {} + \lambda(w\transp (\frac{1}{2}Y)z+z\transp (\frac{1}{2}Y\transp )w-w\transp w)<0
\end{aligned}
\end{equation*}

The condition can be rewritten as~\eqref{eq:dTMd_performance},
which is the LMI
\begin{equation*}
    \mathcal{L}(Q,Y,\mu)=\begin{bmatrix}
        \mathcal{X} & \mathcal{\widetilde{B}}+\frac{1}{2}\lambda\mathcal{\widetilde{Y}}\transp \\\mathcal{\widetilde{B}}\transp +\frac{1}{2}\lambda\mathcal{\widetilde{Y}} & -\lambda
    \end{bmatrix} <0,
\end{equation*}
where
$\mathcal{X}$ is the upper left block matrix in~\eqref{eq:dTMd_performance},
and
\begin{equation*}
\resizebox{\columnwidth}{!}{$
    \mathcal{\widetilde{B}}\transp = \begin{bmatrix}
        -B\transp & \smash{\underbrace{0_{1\times n} \;\cdots\; 0_{1\times n}}_{m \text{ times}}}
    \end{bmatrix}, \quad
    \mathcal{\widetilde{Y}}= \begin{bmatrix}
        Y & \smash{\underbrace{0_{1\times n} \;\cdots\; 0_{1\times n}}_{m \text{ times}}}
    \end{bmatrix}.
    \vphantom{\underbrace{0}_{m \text{ times}}}
$}
\end{equation*}

The term \(\lambda \mathcal{\widetilde{Y}}\) introduces a bilinearity, which can be convexified as follows. By applying the Schur complement~\cite{zhang2006schur}, the LMI is negative definite if and only if \(-\lambda<0\) and \(\mathcal{X}-(\mathcal{\widetilde{B}}+\frac{1}{2}\lambda\mathcal{\widetilde{Y}}\transp )(\frac{1}{-\lambda})(\mathcal{\widetilde{B}}\transp +\frac{1}{2}\lambda\mathcal{\widetilde{Y}})<0\). The first condition holds by construction. We can rewrite the second condition as
\begin{equation*}
\begin{aligned}
\mathcal{X} + \tfrac{1}{\lambda}\mathcal{\widetilde{B}}\mathcal{\widetilde{B}}\transp 
+ \tfrac{1}{2}\Big(\mathcal{\widetilde{B}\widetilde{Y}} + \mathcal{\widetilde{Y}}\transp \mathcal{\widetilde{B}}\transp \Big)
- \mathcal{\widetilde{Y}}\transp \Big(\tfrac{1}{\frac{-4}{\lambda}}\Big)\mathcal{\widetilde{Y}}<0,
\end{aligned}
\end{equation*}
which, using the Schur complement, can be recast into the LMI
\begin{equation}\label{eq:LMI31}
    \begin{bmatrix}
        \mathcal{X}+\frac{1}{\lambda}\mathcal{\widetilde{B}}\mathcal{\widetilde{B}}\transp +\frac{1}{2}(\mathcal{\widetilde{B}\widetilde{Y}}+\mathcal{\widetilde{Y}}\transp \mathcal{\widetilde{B}}\transp ) & \mathcal{\widetilde{Y}}\transp  \\ \mathcal{\widetilde{Y}} & \frac{-4}{\lambda}
    \end{bmatrix}<0,
\end{equation}
which is convex in the decision variables \(Q\), \(Y\), \(\frac{1}{\lambda}\), and \(\mu_i\).

Since global exponential stability cannot be ensured from a bounded input, and global asymptotic stability requires that the plant matrix \(A(d)\) is not exponentially unstable (which is not guaranteed), the control input saturation value can be enforced by imposing a constraint on the linear region of the control law, together with the usual region-of-attraction constraint~\cite{da2005antiwindup}. We do this by expressing the sector condition as
\begin{equation*}
    \phi (Kx)\transp (Kx-\phi (Kx)+Hx)\geq0.
\end{equation*}
We also require the stability to be regionally exponential:
\begin{equation}\label{ineq_regional_stab}
    V(x)=x\transp Px\leq x(0)\transp Px(0)\leq R^2\implies \Vert Hx\Vert \leq u_{\max}^2,
\end{equation}
which can be rewritten as an LMI:
\begin{equation*}
    V(x)=x\transp Px\leq R^2 \implies x\transp (u_{\max}^2\frac{P}{R^2})x\leq u_{\max}^2
\end{equation*}
We can ensure Inequality~\eqref{ineq_regional_stab} as follows:
\begin{equation*}
\begin{aligned}
x\transp H\transp Hx < x\transp \Big(u_{\max}^2 \tfrac{P}{R^2}\Big)x
\iff {} &
\begin{bmatrix}
P & H\transp  \\
H & \tfrac{u_{\max}^2}{R^2}
\end{bmatrix} > 0
\\
\overset{Y_H := HQ}{\iff} &
\begin{bmatrix}
        Q & Y_H\transp \\ Y_H &\frac{u_{\max}^2}{R^2}
    \end{bmatrix}>0
\end{aligned}
\end{equation*}
With \(\mathcal{\widetilde{Y}}_H = \begin{bmatrix}
        Y_H & \smash{\underbrace{0_{1\times n} \quad \cdots \quad 0_{1\times n}}_{m \text{ times}}}
    \end{bmatrix}
    \vphantom{\underbrace{0}_{m \text{ times}}}\), the zonotopic LMI is
\begin{equation}\label{eq:LMI32}
    \begin{bmatrix}
        \mathcal{X}+\frac{1}{\lambda}\mathcal{\widetilde{B}}\mathcal{\widetilde{B}}\transp +\frac{1}{2}(\mathcal{\widetilde{B}\widetilde{Y}}+\mathcal{\widetilde{Y}}\transp \mathcal{\widetilde{B}}\transp ) & \mathcal{\widetilde{Y}}\transp +\mathcal{\widetilde{Y}_H}\transp  \\ \mathcal{\widetilde{Y}}+\mathcal{\widetilde{Y}_H} & \frac{-4}{\lambda}
    \end{bmatrix}<0
\end{equation}

\subsubsection{Decay rate}

We can extend the zonotopic LMI formulation in the usual way to respect a decay rate (\(\alpha_{\text{decay}}>0\)):
\begin{equation*}
    \dot{V}(x)+\alpha_{\text{decay}}V(x)\leq0, \quad \forall x \neq 0
\end{equation*}
The zonotopic LMI is derived following Section~\ref{sec:zono_frame}.

\subsection{Complexity Comparison Between the Zonotopic Approach and the Vertex-Based Approach}

We reformulated the control problem into a single LMI. This structure fundamentally avoids the need for vertex enumeration, which requires defining and solving \(2^m\) distinct LMIs (where $m$ is the number of uncertain/varying parameters; assuming independent intervals for each parameter). Consequently, our formulation only requires the evaluation of system matrices at the center of the zonotope, significantly simplifying the setup phase of the controller design. While the new LMI has a higher dimension, saving the exponential factor reduces the complexity to a polynomial in $m$ (\(\sim \mathcal{O}(m^3)\)).

To evaluate the scalability of our proposed approach, we conducted a comparison against the vertex-based method, focusing only on computational efficiency (synthesis time and memory usage) as the number of varying parameters/uncertainties increases. The test system consists of a 4th-order open-loop unstable plant \(A_0\in\mathbb{R}^{4\text{x}4}\), and to ensure a rigorous and meaningful stress test, the affine matrices \(A_i\) corresponding to each parameter \(\rho_i\in[0,1]\) (up to 20 parameters) were chosen to be sparse but significant in a way that affects different dynamic couplings between states to ensure the parameter dependence matters, and maximizing the computational difficulty for the LMI solver as the number of parameters grows.
\begin{figure}[t]
    \centering
    \includegraphics[width=\linewidth]{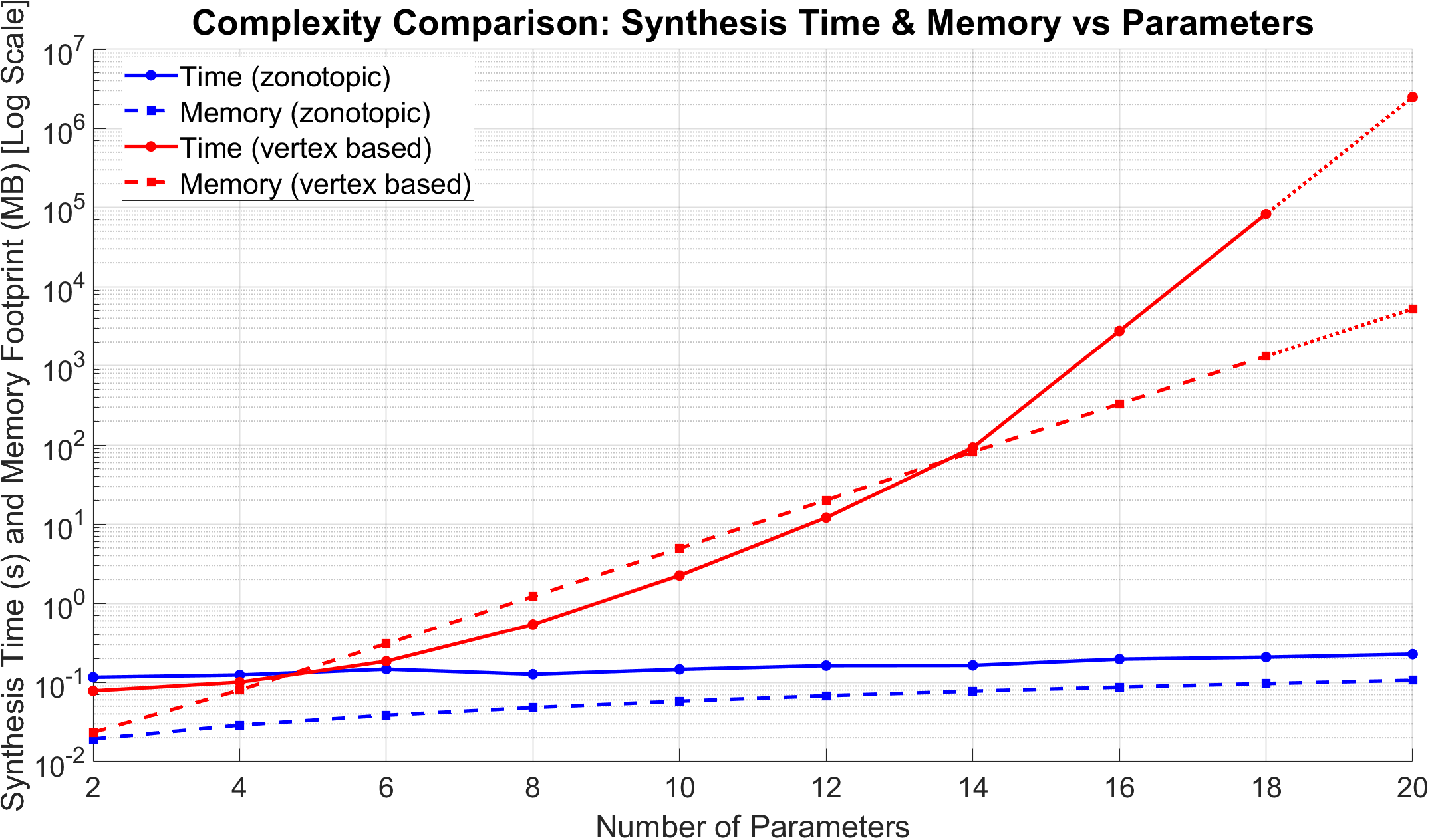}
    \caption{Complexity comparison between the proposed approach vs. the vertex-based approach: Synthesis time and Memory}
    \label{fig:complexity}
\end{figure}
Figure~\ref{fig:complexity} illustrates the computational scalability of both methods: the vertex-based method exhibits an exponential growth in both time and memory consumption, exceeding the limits of our machine for 20 parameters, while in contrast, the complexity of the zonotopic approach scales very favorably, requiring less than 0.3 seconds and negligible memory even at 20 parameters. To minimize measurement variance, computation times for instances requiring less than 2 minutes were averaged over 100 runs.

\subsection{Discussion}

Our approach only yields a sufficient certificate,
as enforcing a single quadratic certificate over a whole zonotope may rule out controllers that can work in practice. (This limitation however also applies to the vertex-based method). The structure of the zonotope affects conservatism and problem size (more generators require more multipliers and larger LMIs). Another limitation is the eventual appearance of bilinear terms when including performance specifications, which cannot be eliminated; hence, iterative heuristics may be needed. Tight performance specifications can make the LMI infeasible without reducing the zonotope size, and numerical conditioning
may be challenging for large problems.

\section{Experimental Evaluation}
\label{sec:results}
In this section, we evaluate the proposed robust zonotopic state feedback synthesis and compare it to several baselines. Our goals are to: (1)~demonstrate that a single gain synthesized over a matrix zonotope can robustly stabilize a wide family of plants whose dynamics vary inside the zonotope, (2)~quantify the trade-offs in performance and conservatism compared to common alternative designs (specifically: (i)~a single common controller synthesized at the vertices, (ii)~a \(\mu\)-synthesis controller, and (iii)~a nominal \(H_{\infty}\) controller), and (3)~assess how additional design requirements (decay rate, actuator limits and saturation handling, and certified regions of attraction) affect the feasibility and performance of the closed-loop system.

First, we perform all tests on the following 4-state example (inspired by two coupled mass-spring-damper systems):
\begin{equation*}
\resizebox{\columnwidth}{!}{$
    \dot{x} = A(d)x + Bu,~
    A(d) = A_0 + d_1A_1 + d_2A_2 + d_3A_3,~
    d_i \in [-1,1]
$}
\end{equation*}
\begin{equation*}
\begingroup
\setlength{\arraycolsep}{0.8mm}
\makebox[\columnwidth][l]{%
\resizebox{\columnwidth}{!}{$
\begin{aligned}
A_0 &=
\begin{bmatrix}
    0 & 1 & 0 & 0 \\
    -3.24 & 0.072 & 0 & 0 \\
    0 & 0 & 0 & 1 \\
    0 & 0 & -12.25 & -0.07
\end{bmatrix}\!,
~
A_1 =
\begin{bmatrix}
    0 & 0 & 0 & 0 \\
    -1.4 & 0.0525 & 0 & 0 \\
    0 & 0 & 0 & 0 \\
    0 & 0 & -2.1 & -0.07
\end{bmatrix}\!,
\\[6pt]
A_2 &=
\begin{bmatrix}
    0 & 0 & 0 & 0 \\
    0 & 0 & 1.05 & 0 \\
    0 & 0 & 0 & 0 \\
    -0.7 & 0 & 0 & 0
\end{bmatrix}\!,
~
A_3 =
\begin{bmatrix}
    0 & 0 & 0 & 0 \\
    0 & -0.0875 & 0 & 0.035 \\
    0 & 0 & 0 & 0 \\
    0 & 0.0175 & 0 & -0.14
\end{bmatrix}\!,
~
B =
\begin{bmatrix}
    0 \\
    1 \\
    0 \\
    0.5
\end{bmatrix}
\end{aligned}
$}}
\endgroup
\end{equation*}

The open-loop dynamics is unstable for a substantial part of the ($3$-dimensional) parameter domain, so this system provides an explanatory stress test for robust synthesis methods.
For fair comparisons, we synthesize the following controllers: (1)~the proposed zonotopic single static gain \(K_{\text{zono}}\) (one gain for the whole zonotope), (2)~a vertex common-$K$ controller \(K_{\text{poly}}\) obtained by enforcing a single gain for stability at every vertex (\(2^3\) vertices) of the parameter domain, (3)~a \(\mu\)-controller \(K_{\mu}\) (used as a robust-performance reference, although it is not a single gain like \(K_{\text{zono}}\) and~\(K_{\text{poly}}\)), and (4)~a nominal \(H_{\infty}\) controller \(K_{\text{Hinf}}\) designed on \(A_0\). We synthesize \(K_{\mu}\) ($D$-$K$ iteration) given the uncertain plant \(A\) (by specifying the uncertain elements of the matrix~$A$ by nominal value and range of uncertainty). We design \(K_{\text{Hinf}}\) on the nominal model \(A_0\) and then test it at every vertex; its robustness is also analyzed across the entire zonotope.

We evaluate the controllers using robust stability tests under actuator constraints (added through a sector-bounded saturation): we report the synthesis feasibility and examine the stability coverage and time-domain behavior across the parameter domain. We also report the percentage of the sampled plants stabilized, the worst-case eigenvalue, representative time responses, and the control effort. The synthesis is implemented in YALMIP/MATLAB~\cite{yalmip} with SDPT3~\cite{sdpt3} as solver. For \(K_{\text{zono}}\), we solve the LMI formulated as explained in Section~\ref{sec:control_pb}, tailored to our example. For \(K_{\text{poly}}\), the decision variables ($Q$ and $Y$) are the same for every vertex~\(A_v\):
\begin{equation*}
    A_vQ+QA_v\transp +BY+Y\transp B\transp <0
\end{equation*}
The 8 LMIs (size 4) are enforced jointly, and the recovered~$K$ is valid for all vertices and hence also for the whole zonotope. The actuator constraint is added as explained in Section~\ref{sec:sectorbound}: we impose \(u_{\max}=5\) for both \(K_{\text{zono}}\) and \(K_{\text{poly}}\) as a hard constraint and ensure that the initial condition \(x_0\) is inside the regional exponential stability region, i.e., \(x_0\transp Px_0\le R^2\), which can be expressed as an LMI (using Schur complement) as
\begin{equation*}
    \begin{bmatrix}
        \gamma & \gamma x_0\transp \\
        \gamma x_0 & Q
    \end{bmatrix}\geq 0, \quad \text{ where } \gamma=\frac{1}{R^2}>0.
\end{equation*}
Solving the LMIs to minimize \(\gamma\) yields the largest~\(R\) that contains~\(x_0\) and satisfies the robust stability and the sector non-linearity constraint. Actuator saturation can only be added as a soft constraint to \(K_{\mu}\) and \(K_{\text{Hinf}}\) via the weighting~\(W_u = \frac{1}{u_{\max}}=0.2\). We do not use any anti-windup scheme, and test the hard saturation constraint during the simulations.

\begin{figure}[t]
    \centering
    \includegraphics[width=\linewidth]{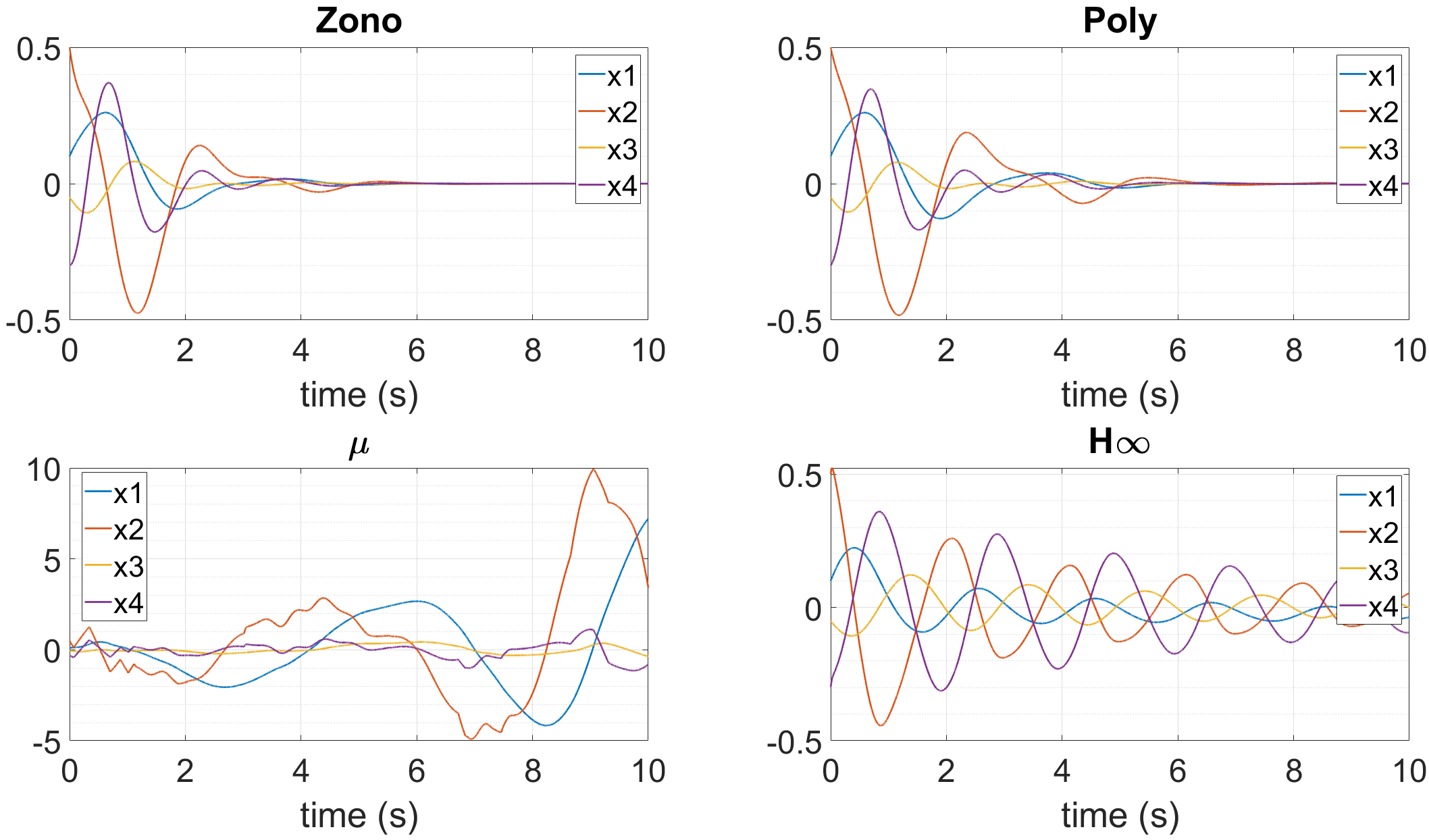}
    \caption{Time domain responses for case~\ref{it:sinusoid} (performance under input constraints): state trajectories}
    \label{fig:states case 2 - sat}
\end{figure}

\begin{figure}[t]
    \centering
    \includegraphics[width=0.89\linewidth]{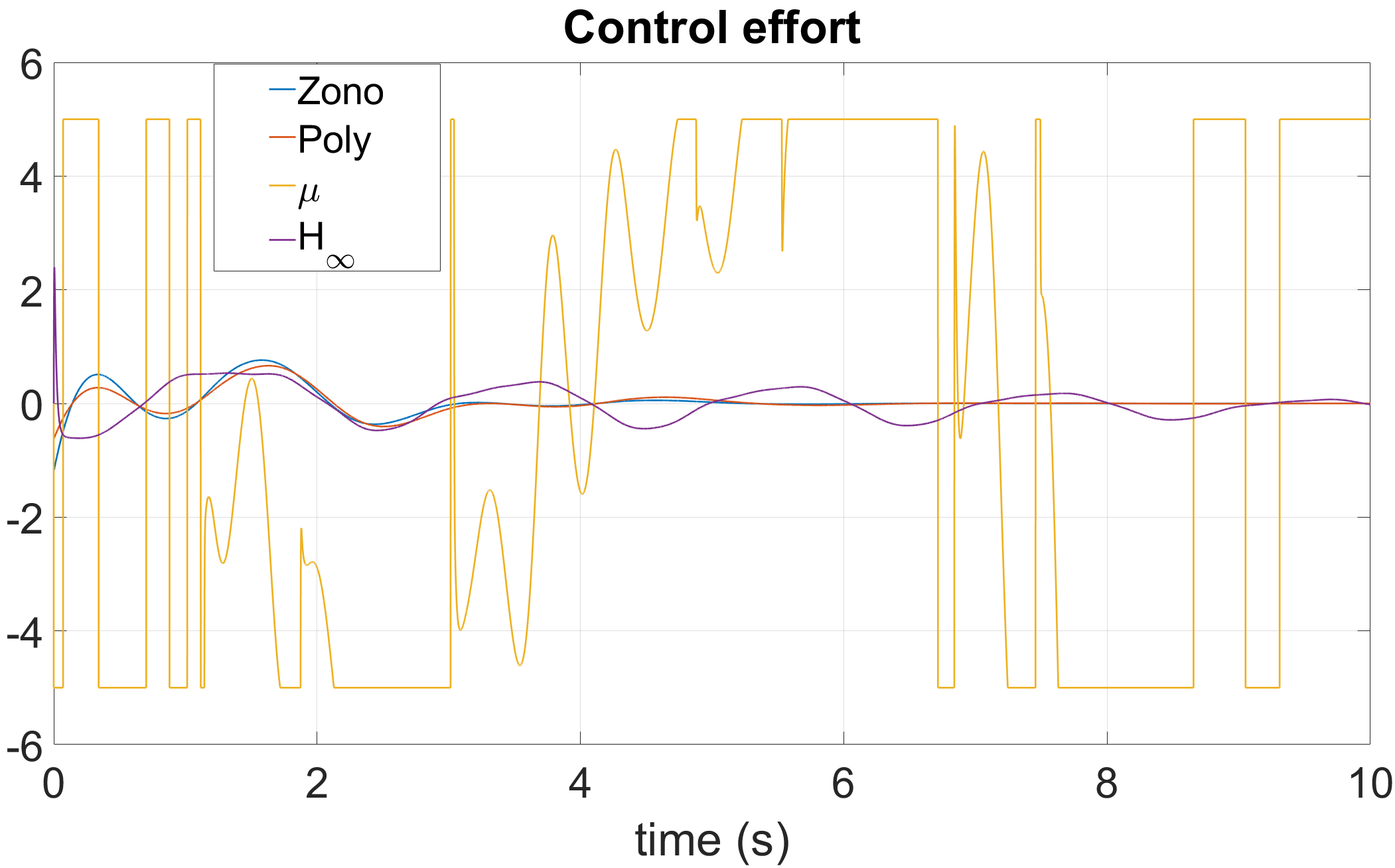}
    \caption{Control input for case~\ref{it:sinusoid} (performance under input constraints)}
    \label{fig:control case 2 - sat}
\end{figure}

\begin{figure}[t]
    \centering
    \includegraphics[width=\linewidth]{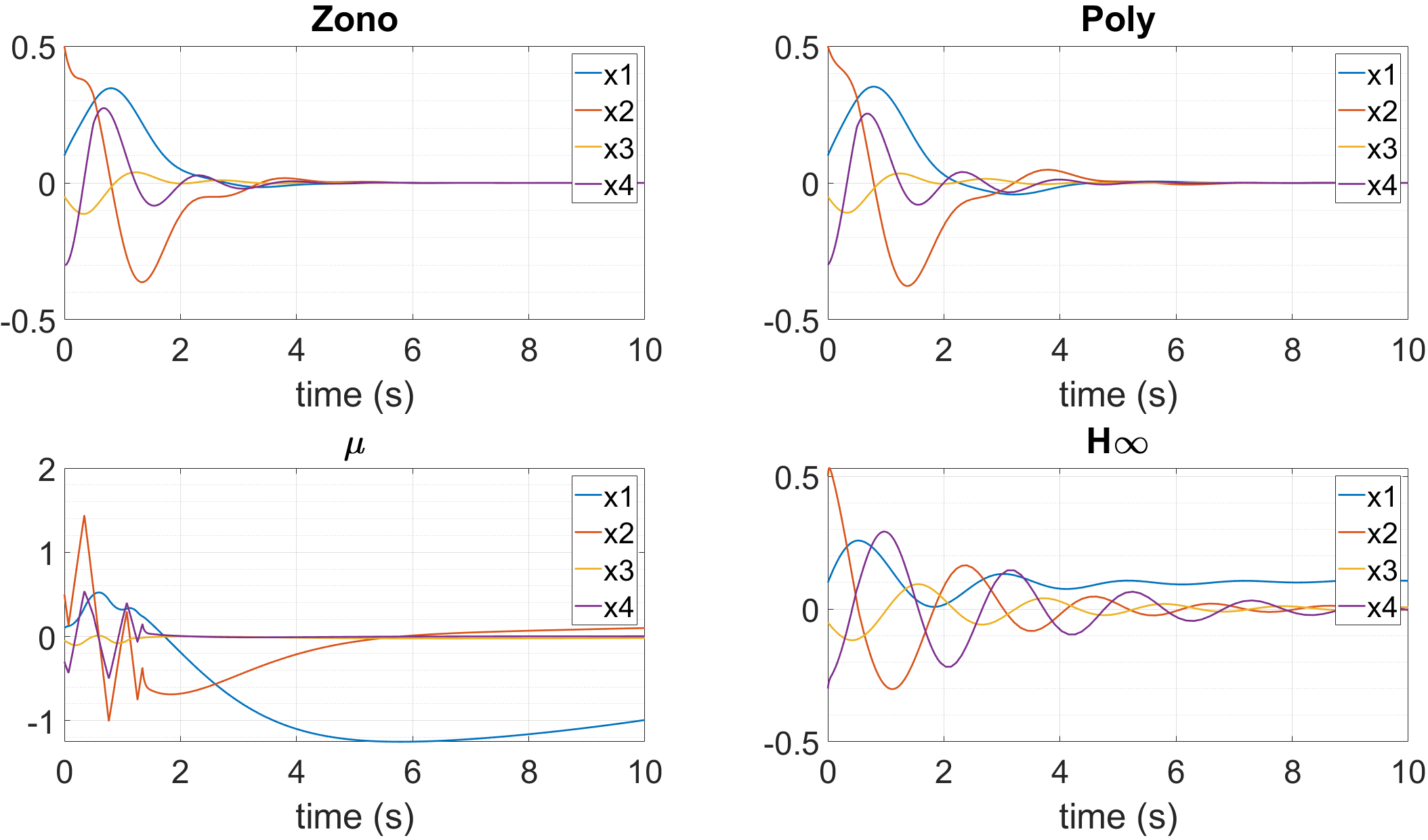}
    \caption{Time domain responses for case~\ref{it:abrupt} (performance under input constraints): state trajectories}
    \label{fig:states case 3 - sat}
\end{figure}

\begin{figure}[t]
    \centering
    \includegraphics[width=0.89\linewidth]{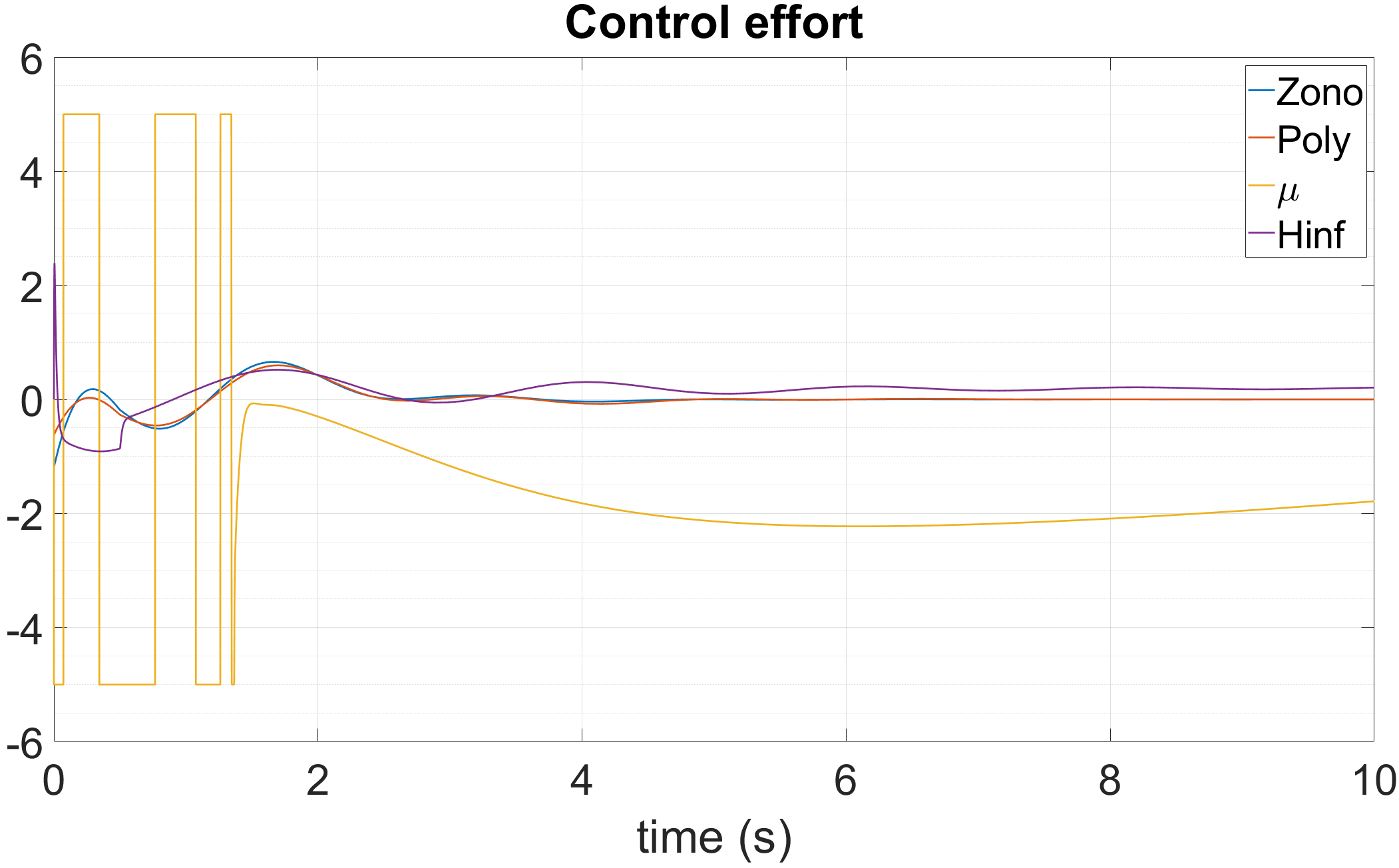}
    \caption{Control input for case~\ref{it:abrupt} (performance under input constraints)}
    \label{fig:control case 3 - sat}
\end{figure}

\subsubsection{Synthesis summary}
\textbf{(i)~Feasibility:} All approaches were feasible. The static gains returned are the following:

\(K_{zono}=\begin{bmatrix}
    -2.150 & -4.123 & -19.554 & -0.363
\end{bmatrix}\),

\(K_{poly}=\begin{bmatrix}
    -1.592 & -2.749 & -13.509 & -0.782
\end{bmatrix}\).
For \(K_{\mu}\), we obtained a dynamic controller of order~58, which we could reduce to order~7 (while retaining the performance of the original \(\mu\)-synthesis controller). For \(K_{\text{Hinf}}\), we obtained a controller of order 4.
\textbf{(ii)~Vertex stability:} The maximum real part of the closed-loop eigenvalues (across all 8 vertices) are -0.919, -0.674, -0.108, 0.0229 for \(K_{\text{zono}}\), \(K_{\text{poly}}\), \(K_{\mu}\), and \(K_{\text{Hinf}}\), respectively, confirming stability for the first two at every vertex, as well as for \(K_{\mu}\) for the non-saturated case. Appendix~\ref{apptimevertexactconst1} shows transients and control efforts (for all vertices). Under the actuator constraint (\(|u_{\max}|\leq5\)), \(K_{\text{zono}}\) ensures stability and a performance similar to \(K_{\text{poly}}\). The \(K_{\mu}\) controller fails to stabilize the plant for vertices 1, 2, 4, and 8 due to saturation, indicating need for anti-windup. The \(K_{\text{Hinf}}\) controller cannot stabilize the plant for vertices 2, 3, and~7.
\textbf{(iii)~Robustness:} Having the respective LMIs for \(K_{\text{zono}}\) and \(K_{\text{poly}}\) feasible (under the actuator constraint LMI) proves their robustness across the whole parameter domain, and \(K_{\mu}\) achieved a robustness up to 223\% of this parameter domain, whereas \(K_{\text{Hinf}}\) only achieved up to 88.8\%. In a Monte Carlo test on \(100{,}000\) sample plants, \(K_{\text{zono}}\) and \(K_{\text{poly}}\) were stable in \(100\%\) of the tested plants, whereas \(K_{\mu}\) and \(K_{\text{Hinf}}\) were stable in only \(51.5\%\) and \(67.7\%\), respectively.

\subsubsection{Time domain evaluation}
\label{sec:time_do_1}
We validate the closed-loop behavior in the time domain to complement the vertex stability check and sampling tests. The goal is to demonstrate transient performance, robustness to parameter variation, and relative control effort under varying operating conditions.

\textbf{Simulation setup:} All time-domain tests are run in continuous time (MATLAB ode15s)
with time horizon \(10s\).

\textbf{Parameter cases:}
We test three classes of parameters \(d_i\):
\begin{enumerate}
    \item \label{it:vertices} Constant values \(d_i\) at the domain vertices, which often represent the worst-case instances and thus challenging fixed plants. (For time-domain graphs, see appendix~\ref{apptimevertexactconst1}.)
    \item \label{it:sinusoid} Fast-varying parameters, where each \(d_i\) follows an independent frequency sinusoid in $[-1,1]$
    ($f_1 = 0.5$, $f_2 = 0.2$, $f_3 = 2$; in Hz) with phases $\varphi_1 = 0$, $\varphi_2 = 1.131$, $\varphi_3 = 2.3$) (in rad, see Appendix~\ref{app d_i case 2 case 3}).  We also perform a high-frequency stress test (\(f_1 = f_2 = f_3 = 5\) Hz).
    \item \label{it:abrupt} Abrupt switch between two opposite vertices to stress transient robustness.
\end{enumerate}

\textbf{Initial conditions:} We use \(x_0\transp = [0.1, 0.5, -0.05, -0.3]\), drawn near an eigenvector of an open-loop unstable plant.

The transient behavior and controller effort are shown in figures~\ref{fig:states case 2 - sat}, \ref{fig:control case 2 - sat}, \ref{fig:states case 3 - sat}, and~\ref{fig:control case 3 - sat}.

With control saturation, \(K_{\mu}\) fails to stabilize the plant under fast parameter variations, which is expected, as it is designed for static or slowly-varying uncertainties and worst-case scenarios. The \(K_{\text{Hinf}}\) controller in case~2 is stable but much slower than the other controllers, and is unstable in case~3. In contrast, both \(K_{\text{zono}}\) and \(K_{\text{poly}}\) stabilize the plant in all cases with similar transients, and their control effort stays below the imposed limit. Table~\ref{tab:metricsB} summarizes the performance.

These results prove that, even under additional constraints, \(K_{\text{zono}}\) still achieves the desired stability coverage and performances in terms of control effort, exhibiting similar transient performance and control effort to \(K_{\text{poly}}\) even for fast-varying parameters, and could even stabilize the plant under fast-changing operational points with limited control effort, whereas the robust \(\mu\)-controller failed.

\begin{table}[t]
    \centering
    \begin{tabular}{ccccc}
       Metric & \(K_{\text{zono}}\) & \(K_{\text{poly}}\) & \(K_{\mu}\) & \(K_{\text{Hinf}}\) \\
       \toprule
       Robust Stability Margin & 2.07 & 2.16 & 2.23 & 0.89\\
       Peak state norm  & 0.326 & 0.326 & unstable & 0.363\\
       RMS state norm  & 0.184 & 0.193 & 2.468 & 0.240\\
       Peak control  & 1.190 & 0.664 & 5 & 2.400\\
       RMS control  & 0.220 & 0.188 & 4.368 & 0.298\\
       Controller order  & 0 & 0 & 7 & 4\\
    \end{tabular}
    \caption{Performance metrics for stability + input saturation}
    \label{tab:metricsB}
\end{table}

\section{Conclusion}
\label{sec:conclusion}

We proposed zonotopic robust control, a novel approach to synthesizing a single, robust state feedback controller that certifies stability over a matrix zonotope. Using generator-aware multipliers, we convert the infinite family of Lyapunov inequalities into a single convex LMI. In numerical experiments, we demonstrated that our approach often achieves larger stability coverage across the parameter domain while keeping synthesis complexity low, providing a useful, computationally efficient alternative to vertex enumeration and \(\mu\)-synthesis, and a practical path toward less conservative single-gain robust controllers. Overall, our approach is a promising way to exploit zonotopic structure for robust and adaptive control design and synthesis. Future works will focus on extending the approach into a self-scheduled adaptive control via the usage of parameter-dependent Lyapunov functions to reduce conservatism, incorporating advanced specifications such as \(H_{\infty}\) performance, and performing experimental validation.

\section*{acknowledgment}
This research was partly supported by the Independent Research Fund Denmark under reference number 10.46540/3120-00041B and the Villum Investigator Grant S4OS under reference number 37819.

\bibliographystyle{IEEEtran}
\bibliography{bibliography}

@inproceedings{AlthoffGK11,
  author       = {Matthias Althoff and
                  Colas Le Guernic and
                  Bruce H. Krogh},
  editor       = {Marco Caccamo and
                  Emilio Frazzoli and
                  Radu Grosu},
  title        = {Reachable set computation for uncertain time-varying linear systems},
  booktitle    = {{HSCC}},
  pages        = {93--102},
  publisher    = {{ACM}},
  year         = {2011},
  @url          = {https://doi.org/10.1145/1967701.1967717},
  doi          = {10.1145/1967701.1967717}
}

@book{zhou_lpv,
  title={Essentials of robust control},
  author={Zhou, Kemin and Doyle, John Comstock},
  volume={104},
  year={1998},
  publisher={Prentice hall Upper Saddle River, NJ}
}

@book{toth2010modeling,
  title={Modeling and identification of linear parameter-varying systems},
  author={T{\'o}th, Roland},
  volume={403},
  year={2010},
  publisher={Springer}
}

@article{wu1996induced,
  title={Induced {L2}-norm control for {LPV} systems with bounded parameter variation rates},
  author={Wu, Fen and Yang, Xin Hua and Packard, Andy and Becker, Greg},
  journal={International Journal of Robust and Nonlinear Control},
  volume={6},
  number={9-10},
  pages={983--998},
  year={1996},
  publisher={Wiley Online Library}
}

@book{mohammadpour2012control,
  title={Control of linear parameter varying systems with applications},
  author={Mohammadpour, Javad and Scherer, Carsten W},
  year={2012},
  publisher={Springer Science \& Business Media}
}

@phdthesis{althoff2010reachability,
  title={Reachability analysis and its application to the safety assessment of autonomous cars},
  author={Althoff, Matthias},
  year={2010},
  school={Technische Universit{\"a}t M{\"u}nchen}
}

@inproceedings{girard2005reachability,
  title={Reachability of uncertain linear systems using zonotopes},
  author={Girard, Antoine},
  booktitle={International workshop on hybrid systems: Computation and control},
  pages={291--305},
  year={2005},
  organization={Springer}
}

@book{boyd1994linear,
  title={Linear matrix inequalities in system and control theory},
  author={Boyd, Stephen and El Ghaoui, Laurent and Feron, Eric and Balakrishnan, Venkataramanan},
  year={1994},
  publisher={SIAM}
}

@article{APKARIAN19951251,
title = {Self-scheduled {H}$\infty$ control of linear parameter-varying systems: a design example},
journal = {Automatica},
volume = {31},
number = {9},
pages = {1251-1261},
year = {1995},
issn = {0005-1098},
doi = {https://doi.org/10.1016/0005-1098(95)00038-X},
@url = {https://www.sciencedirect.com/science/article/pii/000510989500038X},
author = {Pierre Apkarian and Pascal Gahinet and Greg Becker},
keywords = {Scheduled control, robust control, missiles, time-varying systems, convex programming,  control, parametric variations},
abstract = {This paper is concerned with the design of gain-scheduled controllers with guaranteed H∞ performance for a class of linear parameter-varying (LPV) plants. Here the plant state-space matrices are assumed to depend affinely on a vector θ of time-varying real parameters. Assuming real-time measurement of these parameters, they can be fed to the controller to optimize the performance and robustness of the closed-loop system. The resulting controller is time-varying and automatically ‘gain-scheduled’ along parameter trajectories. Based on the notion of quadratic H∞ performance, solvability conditions are obtained for continuous- and discrete-time systems. In both cases the synthesis problem reduces to solving a system of linear matrix inequalities (LMIs). The main benefit of this approach is to bypass most difficulties associated with more classical schemes such as gain-interpolation or gain-scheduling techniques. The methodology presented in this paper is applied to the gain scheduling of a missile autopilot. The missile has a large operating range and high angles of attack. The difficulty of the problem is reinforced by tight performance requirements as well as the presence of flexible modes that limit the control bandwidth.}
}

@article{chilali2002h,
  title={{H}$\infty$ design with pole placement constraints: an {LMI} approach},
  author={Chilali, Mahmoud and Gahinet, Pascal},
  journal={IEEE Transactions on automatic control},
  volume={41},
  number={3},
  pages={358--367},
  year={2002},
  publisher={IEEE}
}

@article{yakubovich1971sprocedure,
  author    = {V. A. Yakubovich},
  title     = {S-procedure in nonlinear control theory},
  journal   = {Vestnik Leningrad University},
  volume    = {1},
  pages     = {62--77},
  year      = {1971}
}

@book{sename2025linear,
  title={Linear Parameter-Varying Control: Theory and Application to Automotive Systems},
  author={Sename, Olivier},
  year={2025},
  publisher={John Wiley \& Sons}
}

@article{Set_based_FDI_LPV,
title = "Set-based fault detection and isolation for detectable linear parameter-varying systems",
abstract = "In the context of fault detection and isolation of linear parameter-varying systems, a challenging task appears when the dynamics and the available measurements render the model unobservable, which invalidates the use of standard set-valued observers. Two results are obtained in this paper, namely, using a left-coprime factorization, one can achieve set-valued estimates with ultimately bounded hyper-volume and convergence dependent on the slowest unobservable mode; and by rewriting the set-valued observer equations and taking advantage of a coprime factorization, it is possible to have a low-complexity fault detection and isolation method. Performance is assessed through simulation, illustrating, in particular, the detection time for various types of faults.",
keywords = "coprime factorization, distributed, fault detection and isolation, unobservable LPV",
author = "Daniel Silvestre and Paulo Rosa and Hespanha, \{Jo{\~a}o P.\} and Carlos Silvestre",
@note = "Publisher Copyright: Copyright {\textcopyright} 2017 John Wiley \& Sons, Ltd.",
year = "2017",
month = dec,
day = "1",
doi = "10.1002/rnc.3814",
language = "English",
volume = "27",
pages = "4381--4397",
journal = "International Journal of Robust and Nonlinear Control",
issn = "1049-8923",
publisher = "John Wiley and Sons Ltd",
number = "18",
}

@article{Zono_Set_Switched_LPV,
title = {Zonotopic Set-Membership State Estimation for Switched {LPV} Systems},
journal = {IFAC-PapersOnLine},
volume = {56},
number = {2},
pages = {9442-9447},
year = {2023},
note = {22nd IFAC World Congress},
issn = {2405-8963},
doi = {https://doi.org/10.1016/j.ifacol.2023.10.238},
@url = {https://www.sciencedirect.com/science/article/pii/S240589632300589X},
author = {Shuang Zhang and Vicenc Puig and Sara Ifqir},
keywords = {Set-membership estimation, state estimation, switched polytopic LPV system, parameter uncertainty, average dwell time, zonotopes},
abstract = {This paper addresses the state estimation problem for switched discrete-time Linear Parameter Varying (LPV) systems with mensurable and unmeasurable scheduling parameters. A zonotopic switched polytopic state estimator, considering parameter uncertainty, is proposed based on a Set-Membership Approach (SMA). Taking Average Dwell Time (ADT) into account, a new criterion is proposed to guarantee the convergence of the estimation. An application to vehicle lateral dynamics state estimation is used as case study. Simulation results reveal the effectiveness of the proposed algorithm and demonstrate advantages over the existing methods.}
}

@incollection{althoff2011analyzing,
  title={Analyzing reachability of linear dynamic systems with parametric uncertainties},
  author={Althoff, Matthias and Krogh, Bruce H and Stursberg, Olaf},
  booktitle={Modeling, Design, and Simulation of Systems with Uncertainties},
  pages={69--94},
  year={2011},
  publisher={Springer}
}

@inproceedings{luo2023reachability,
  title={Reachability analysis for linear systems with uncertain parameters using polynomial zonotopes},
  author={Luo, Ertai and Kochdumper, Niklas and Bak, Stanley},
  booktitle={Proceedings of the 26th ACM International Conference on Hybrid Systems: Computation and Control},
  pages={1--12},
  year={2023}
}

@inproceedings{alanwar2021data,
  title={Data-driven reachability analysis using matrix zonotopes},
  author={Alanwar, Amr and Koch, Anne and Allg{\"o}wer, Frank and Johansson, Karl Henrik},
  booktitle={Learning for Dynamics and Control},
  pages={163--175},
  year={2021},
  organization={PMLR}
}

@inproceedings{le2011robust,
  title={Robust tube-based constrained predictive control via zonotopic set-membership estimation},
  author={Le, Vu Tuan Hieu and Stoica, Cristina and Dumur, Didier and Alamo, Teodoro and Camacho, Eduardo F},
  booktitle={2011 50th IEEE Conference on Decision and Control and European Control Conference},
  pages={4580--4585},
  year={2011},
  organization={IEEE}
}

@article{le2013zonotopic,
  title={Zonotopic guaranteed state estimation for uncertain systems},
  author={Le, Vu Tuan Hieu and Stoica, Cristina and Alamo, Teodoro and Camacho, Eduardo F and Dumur, Didier},
  journal={Automatica},
  volume={49},
  number={11},
  pages={3418--3424},
  year={2013},
  publisher={Elsevier}
}

@article{andrade2024tube,
  title={Tube-based model predictive control based on constrained zonotopes},
  author={Andrade, Richard and Normey-Rico, Julio E and Raffo, Guilherme V},
  journal={IEEE Access},
  volume={12},
  pages={50100--50113},
  year={2024},
  publisher={IEEE}
}

@article{zhang2024linear,
  title={Linear quadratic zonotopic control of switched systems: Application to autonomous vehicle path-tracking},
  author={Zhang, Shuang and Ifqir, Sara and Puig, Vicen{\c{c}}},
  journal={IEEE Control Systems Letters},
  volume={8},
  pages={1895--1900},
  year={2024},
  publisher={IEEE}
}

@article{carrizosa2023zonotopic,
  title={Zonotopic-tube-based {LPV} motion planner for safety coordination of autonomous vehicles},
  author={Carrizosa-Rend{\'o}n, {\'A}lvaro and Puig, Vicen{\c{c}} and Nejjari, Fatiha},
  journal={IFAC-PapersOnLine},
  volume={56},
  number={2},
  pages={2220--2225},
  year={2023},
  publisher={Elsevier}
}

@book{SenameBook,
  TITLE = {{Robust Control and Linear Parameter Varying approaches: Application to Vehicle Dynamics}},
  AUTHOR = {Sename, Olivier and Gaspar, Peter and Bokor, Jozsef},
  @URL = {https://hal.science/hal-03655581},
  EDITOR = {Olivier Sename (Ed.) and Peter Gaspar (Ed.) and Jozsef Bokor (Ed.)},
  PUBLISHER = {{Spinger}},
  SERIES = {Lecture Notes in Control and Information Sciences (LNCIS, volume 437)},
  VOLUME = {437},
  PAGES = {397},
  YEAR = {2013},
  MONTH = Feb,
  DOI = {10.1007/978-3-642-36110-4},
  KEYWORDS = {Control ; Linear Parameter Varying Approaches ; Robust Control ; Vehicle Dynamics ; Control},
  HAL_ID = {hal-03655581},
  HAL_VERSION = {v1},
}

@book{khalil2002nonlinear,
  title={Nonlinear systems},
  author={Khalil, Hassan K and Grizzle, Jessy W},
  volume={3},
  year={2002},
  publisher={Prentice hall Upper Saddle River, NJ}
}

@book{zhang2006schur,
  title={The Schur complement and its applications},
  author={Zhang, Fuzhen},
  volume={4},
  year={2006},
  publisher={Springer Science \& Business Media}
}

@article{da2005antiwindup,
  title={Antiwindup design with guaranteed regions of stability: an LMI-based approach},
  author={Da Silva, JM Gomes and Tarbouriech, Sophie},
  journal={IEEE Transactions on Automatic Control},
  volume={50},
  number={1},
  pages={106--111},
  year={2005},
  publisher={IEEE}
}

@inproceedings{yalmip,
address = {Taipei, Taiwan},
author = {L{\"{o}}fberg, J.},
booktitle = {In Proceedings of the CACSD Conference},
title = {{YALMIP} : A Toolbox for Modeling and Optimization in {MATLAB}},
year = {2004}
}

@article{sdpt3,
volume = {11},
author = {Toh, K C and Todd, M J and Tütüncü, R H},
issn = {1055-6788},
journal = {Optimization methods and software.},
number = {1-4},
title = {{SDPT3} — A {M}atlab software package for semidefinite programming, Version 1.3},
year = {1999-01},
}

@book{le2013zonotopes,
  title={Zonotopes: From guaranteed state-estimation to control},
  author={Le, Vu Tuan Hieu and Stoica, Cristina and Alamo, Teodoro and Camacho, Eduardo F and Dumur, Didier},
  year={2013},
  publisher={John Wiley \& Sons}
}

\appendix

\subsection{Time Domain Graphs for Constant Values of \(d_i\) Corresponding to the Vertices of the Zonotope}
\label{apptimevertexactconst1}
\begin{figure}[h!]
    \centering
    \includegraphics[width=0.9\linewidth]{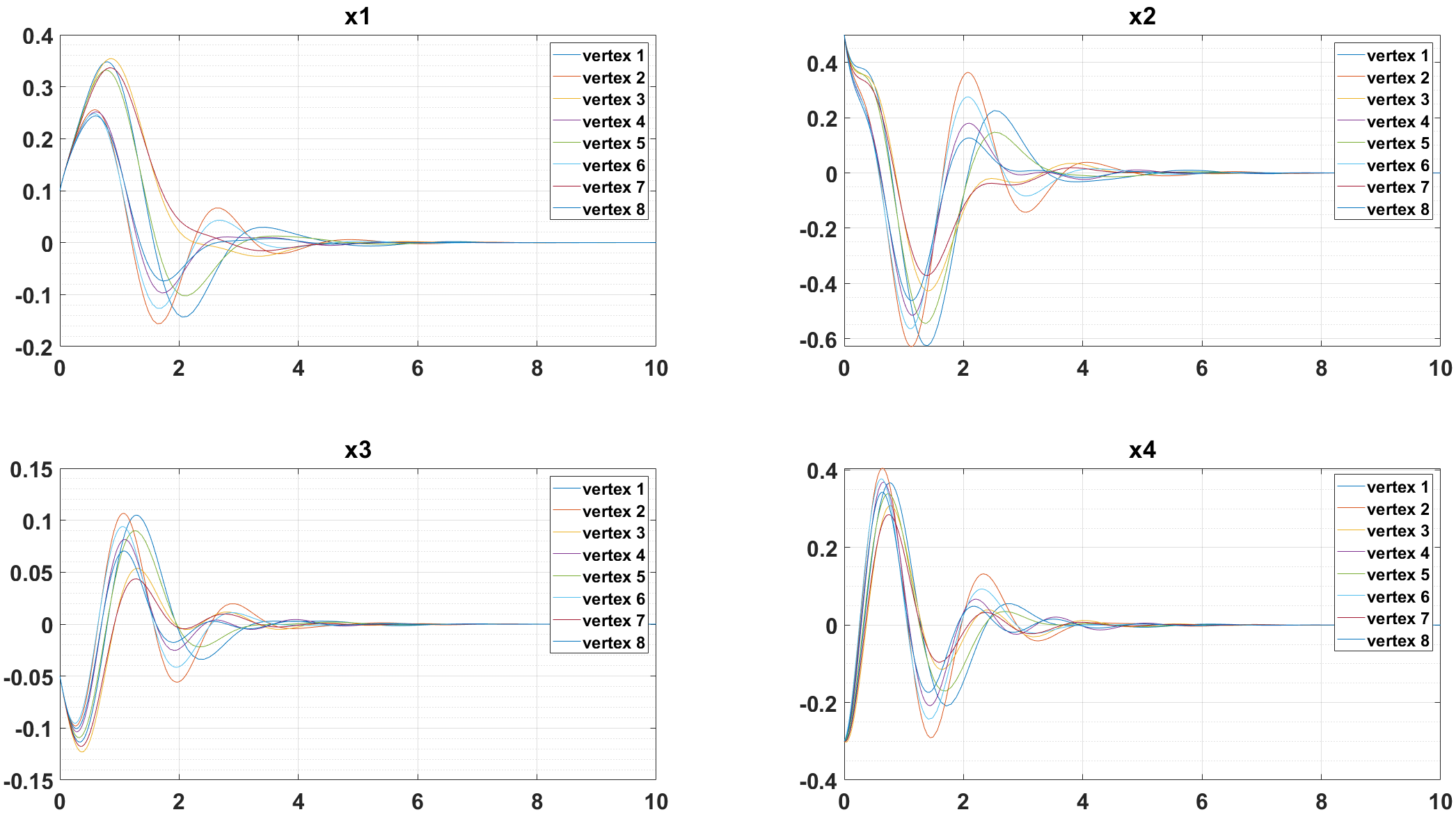}
    \caption{State trajectories for \(K_{\text{zono}}\), actuator constraints}
    \label{fig:zono 1.0}
\end{figure}

\begin{figure}[h!]
    \centering
    \includegraphics[width=0.9\linewidth]{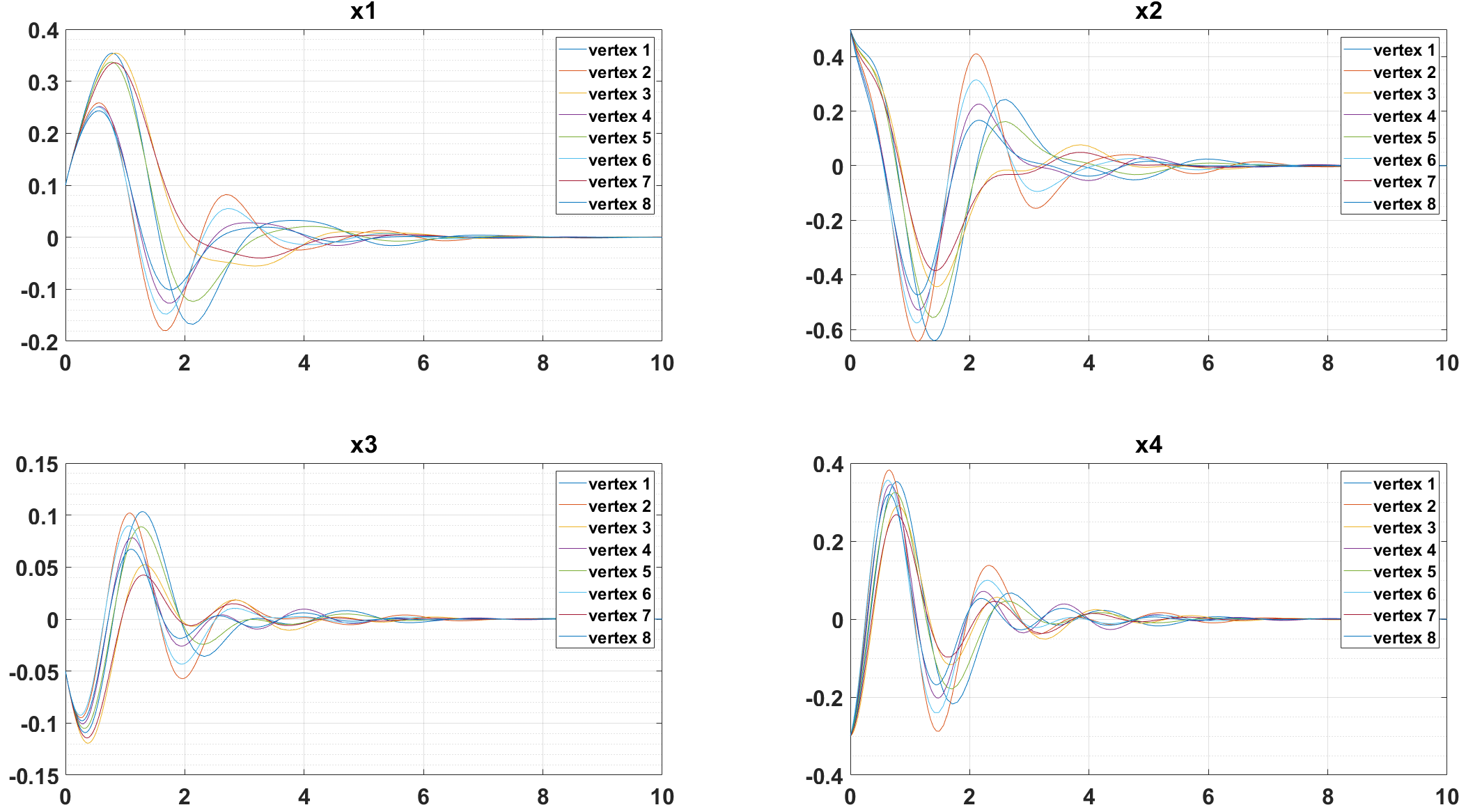}
    \caption{State trajectories for \(K_{\text{poly}}\), actuator constraints}
    \label{fig:poly 1.0}
\end{figure}

\begin{figure}[h!]
    \centering
    \includegraphics[width=0.9\linewidth]{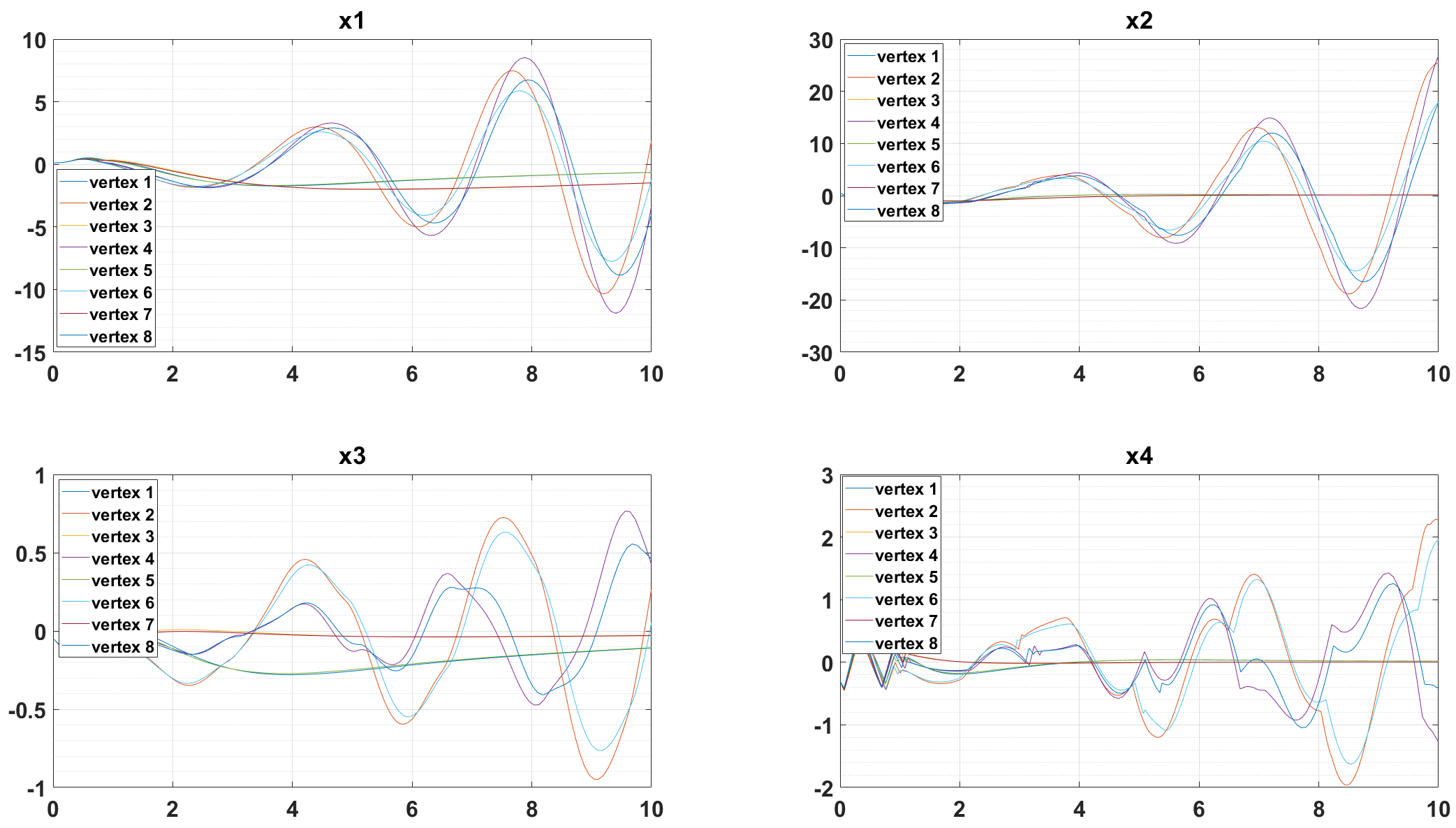}
    \caption{State trajectories for \(K_{\mu}\), actuator constraints}
    \label{fig:mu 1.0}
\end{figure}

\begin{figure}[h!]
    \centering
    \includegraphics[width=0.9\linewidth]{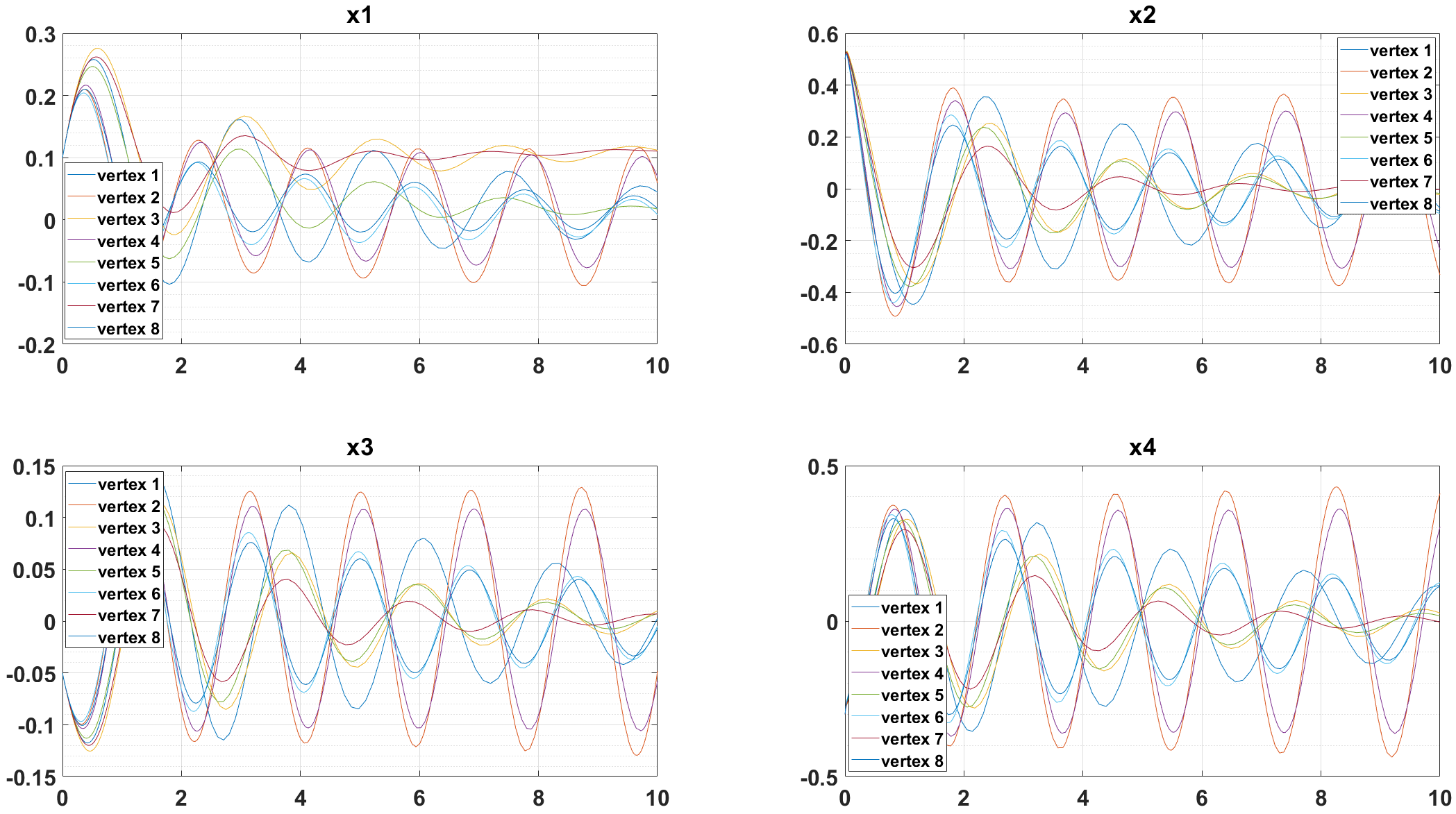}
    \caption{State trajectories for \(K_{\text{Hinf}}\), actuator constraints}
    \label{fig:hinf 1.0}
\end{figure}

\begin{figure}[h!]
    \centering
    \includegraphics[width=0.9\linewidth]{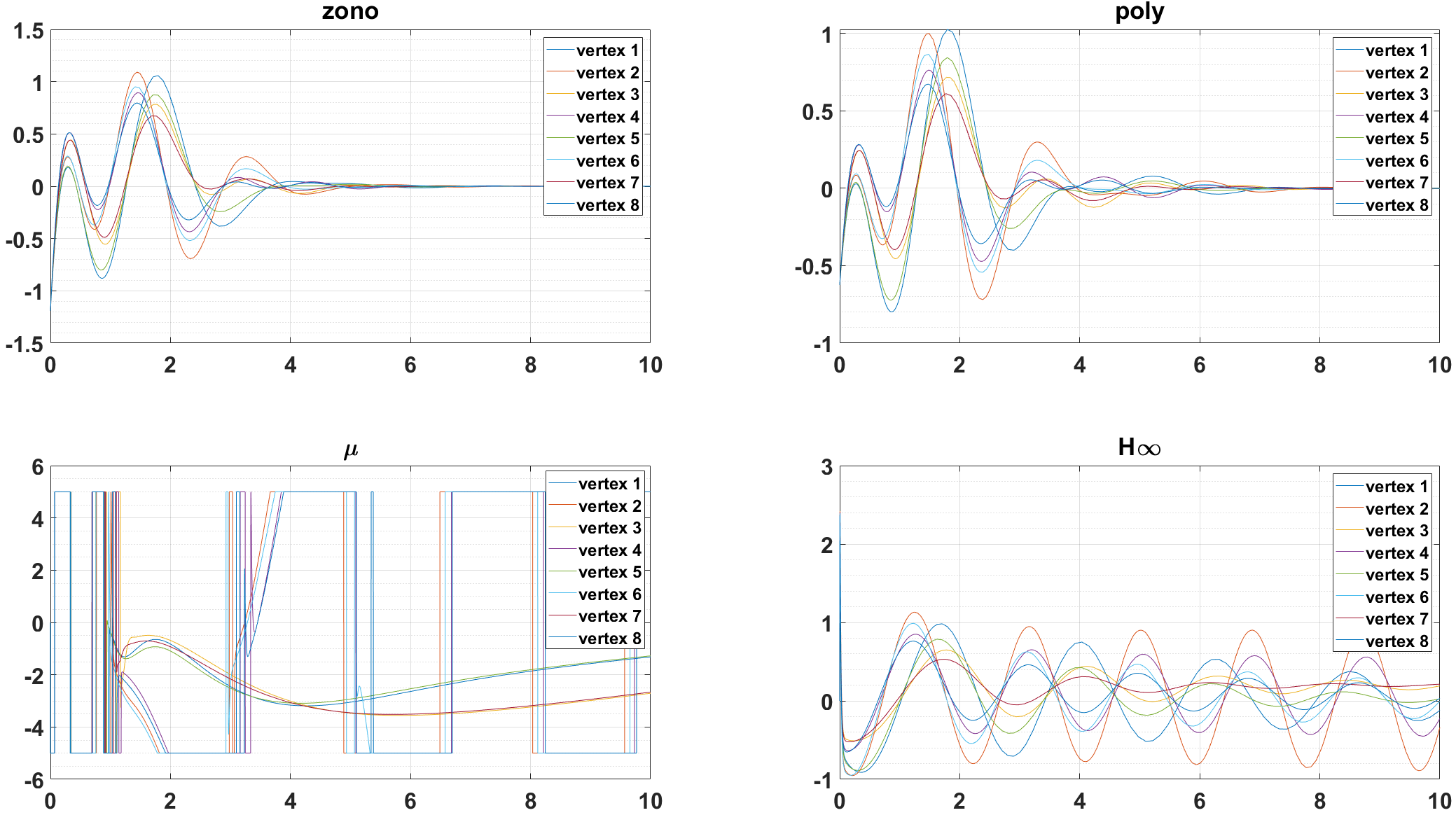}
    \caption{Control effort for all controllers, actuator constraints}
    \label{fig:control 1.0}
\end{figure}

Figures~\ref{fig:zono 1.0} to~\ref{fig:hinf 1.0} show the time-domain graphs of the controllers of Section~\ref{sec:results}, designed and evaluated at each vertex of the considered zonotope, under the specified actuator constraint. \figref{fig:control 1.0} illustrates the corresponding control effort.

\subsection{Time Domain Graphs of the Varying Parameters \(d_i\) for Cases~\ref{it:sinusoid} and~\ref{it:abrupt}}
\label{app d_i case 2 case 3}
\begin{figure}[h!]
    \centering
    \includegraphics[width=\linewidth]{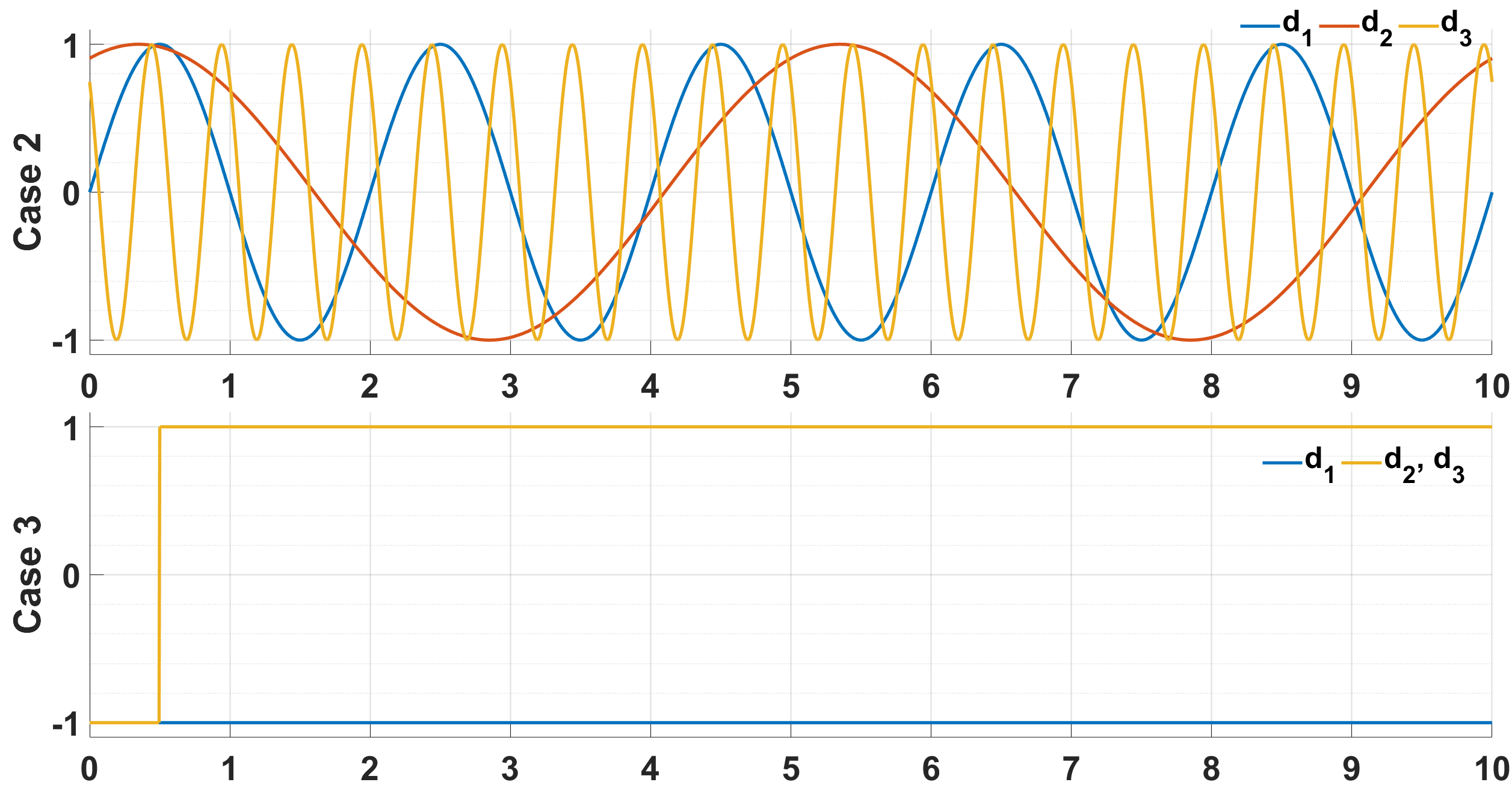}
    \caption{Sequences of \(d_1,d_2,d_3\) for cases~\ref{it:sinusoid} and~\ref{it:abrupt}}
    \label{fig:di case 2}
\end{figure}

Figure \ref{fig:di case 2} shows the sequences of \(d_1\), \(d_2\), and \(d_3\) for cases~\ref{it:sinusoid} and~\ref{it:abrupt} respectively, of Section~\ref{sec:results}.

\end{document}